\documentclass[12pt,preprint]{aastex}
\usepackage{color}

\shortauthors{Pinsonneault \& Stanek}

\shorttitle{Binaries Like to be Twins}

\begin{document}
\title{Binaries Like to be Twins: Implications for Doubly Degenerate
Binaries, the Supernova Ia Rate and Other Interacting Binaries}

\author{M. H. Pinsonneault \& K.~Z.~Stanek}

\affil{Department of Astronomy, The Ohio State University, Columbus, OH 43210}

\email{pinsono@astronomy.ohio-state.edu, kstanek@astronomy.ohio-state.edu}

\begin{abstract}

The recent sample of 21 detached eclipsing binaries in the Small
Magellanic Cloud (Harries et al. 2003, Hilditch et al. 2005) provides
a valuable test of the binary mass function for massive stars.  We
show that 50\% of detached binaries have companions with very similar
masses, $q=M_2/M_1>0.87$, where $M_1,M_2$ denote the masses of the two
binary components, $M_1\ge M_2$.  A Salpeter relative mass function
for the secondary is very strongly excluded, and the data is
consistent with a flat mass function containing 55\% of the systems
and a ``twin'' population with $q>0.95$ containing the remainder.  We
also survey the vast existing literature discussing the mass ratio in
binaries and conclude that a significant twin population (of more than
20-25\%) exists in binaries that are likely to interact across a broad
range of stellar masses and metallicity.  Interactions involving
nearly equal mass stars have distinctly different properties than
those involving stars of unequal mass; the secondaries will tend to be
evolved and the common envelope evolution is qualitatively different.
The implications of such a population for both binary interactions and
star formation are substantial, and we present some examples.  We
argue that twin systems may provide a natural stellar population to
explain the recently proposed prompt channel for type Ia SN, and the
presence of a twin population dramatically reduces the maximum
inferred NS+BH merger rate relative to the NS+NS merger rate.  Twins
may also be important for understanding the tendency of WD and NS
binaries to be nearly equal in mass, and inclusion of twins in
population studies will boost the blue straggler production rate.

\end{abstract}


\section{Introduction}

A majority of stars are in binaries, and a substantial fraction of
binaries have short enough orbital periods that they are likely to
interact during either their main sequence or post-main sequence
evolution.  Many of the most interesting phenomena in astronomy can be
directly traced to the interaction of close binaries; an incomplete
list would include binary neutron stars and white dwarfs, SNIa,
cataclysmic variables, and blue stragglers.  There is a substantial
literature on the subject (e.g. Paczynski 1971). Although there are
many ingredients that must be considered in interacting binaries, an
implicit assumption in much theoretical work has been that the
lifetimes of the stars are almost always quite different.  This arises
naturally from two considerations.  First, the strong mass-lifetime
relationship for all but the most massive stars implies a large
lifetime difference unless the masses are very close.  Second, the
single star IMF is a steep function of mass, with low mass stars being
far more numerous than high mass stars (e.g. Salpeter 1955). If a
similar relative mass function applies to binaries, equal mass systems
are exceedingly unlikely.  Most population synthesis studies adopt a
flat mass spectrum for systems that are likely to interact, following
longstanding evidence that a steeply rising relative IMF is unlikely
(see for example Kuiper 1935).  However, even a flat relative IMF
still yields relatively few systems with very similar masses.
Motivated by recent papers on the production rate of double compact
systems and the possible existence of a prompt channel for type Ia
supernovae (both discussed in Section~4), we investigate the
properties of massive close binaries.

In this paper we present evidence in a sample of massive eclipsing SMC
stars for a substantial population of nearly equal mass binaries.  In
such systems a strong inequality in lifetime is not present, and there
will be important qualitative differences in their evolution compared
to unequal mass binaries.  Furthermore, we argue that recent data in
the literature is entirely consistent with ``twins'' being a general
feature of close binary population.  Our evidence for the presence of
a significant population of twins is presented in Section~2. We
discuss other evidence from the literature in Section~3. We consider
the implications for interacting binary evolution in Section~4.

\section{Sample and Analysis of Massive Binaries in the SMC }

Harries, Hilditch \& Howart (2003; hereafter HHH03) and Hilditch,
Howarth \& Harries (2005; hereafter HHH05) obtained accurate
spectroscopic data to derive a complete set of physical parameters for
50 eclipsing binaries found by the Optical Gravitational Lensing
Experiment in the Small Magellanic Cloud (Udalski et al. 1998;
Wyrzykowski et al. 2004). As discussed by HHH05, their ``full sample
of 50 OB-type eclipsing systems is the largest single set of
fundamental parameters determined for high-mass binaries in any
galaxy''. Of these 50 systems, they find 21 are in detached
configurations, 28 are in semi-detached configurations indicating mass
transfer has occurred, and one is a contact binary. Multi-epoch
spectroscopic data were obtained for a larger sample of 169 systems
(all of them with orbital periods $P<5\;$days), but reliable orbital
solutions were derived for the subset of 50 systems.  The masses
derived for the individual stars in the binaries are typically
uncertain at $\pm 10\%$, with mass of the primary (more massive) star
ranging from $6.9\;M_{\odot}<M_1<27.3\;M_{\odot}$.

\begin{figure}
\plotone{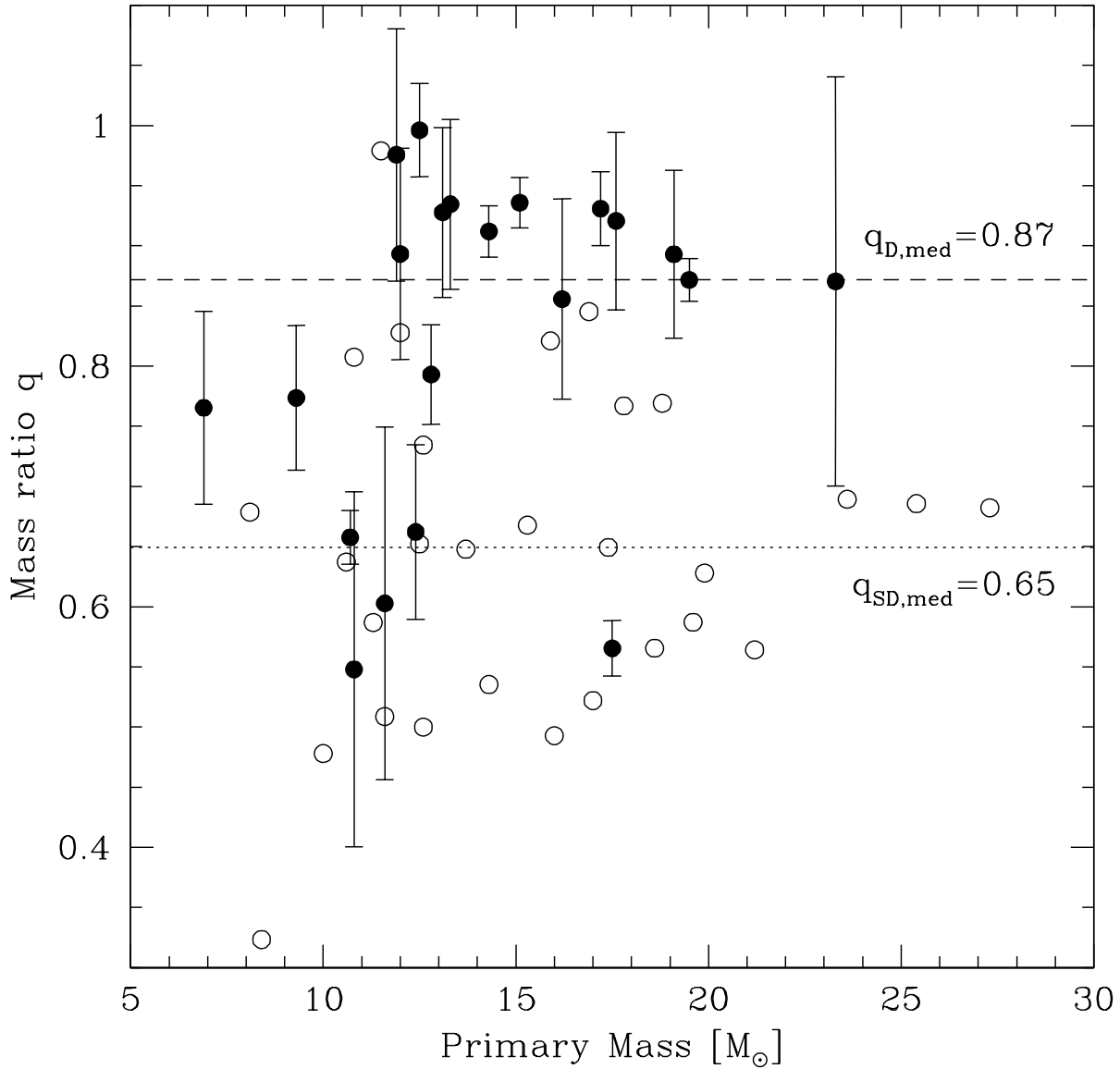}
\caption{Mass ratio $q=M_2/M_1$ for a sample of 50 OB-type eclipsing
binaries from the SMC (Harries et al. 2003; Hilditch et al. 2005)
shown as a function of the primary mass $M_1$. With filled circles and
errorbars we show the sample of 21 detached (unevolved) systems, while
with open circles we show 28 semi-detached and one contact binary.
The median mass ratio is $q_{D,med}=0.87$ for the detached sample and
$q_{SD,med}=0.65$ for the semi-detached/contact sample.}
\label{fig_q}
\end{figure}

The mass ratio $q$ defined as $q=M_2/M_1$, where $M_1\ge M_2$, can be
simply derived from the semi-amplitudes $K_1,K_2$ of the radial
velocity (RV) curves $q=K_1/K_2$, and we show these values for their
full sample of 50 stars in Fig.\ref{fig_q}. With filled circles and
errorbars we show the sample of 21 detached (before any mass transfer
occurred) systems, while with open circles we show 28 semi-detached
and one contact binary.  The median mass ratio is $q=0.87$ for the
detached sample and $q=0.65$ for the semi-detached/contact sample---a
striking difference.  While it is not surprising that the median $q$
value for the semi-detached systems, which underwent mass transfer, is
far from unity, it is very surprising that 13 out of 21 detached
systems have $q>0.85$.  At least at face value, it seems that given
the mass of the primary in a binary, for detached and therefore
unevolved systems in about 50\% of cases the mass of secondary is
within 15\% of the mass of the primary, i.e. they form what we will
call (following Tokovinin 2000 and Halbwachs et al. 2003, see next
Section) ``twins''.

Could that be the result of an observational bias? Clearly, it is
easier to detect eclipsing binaries when the two stars are of similar
size and brightness.  It is also easier to measure the radial
velocities of the two components if they are of similar brightness and
mass.  However, while some biases are certain to be present, it is
very unlikely that the twin population in the SMC sample is a result
of selection effects. There are detached systems present in the sample
with $q$ values as small as 0.55.  In addition, the semi-detached
sample has been selected from the same photometric sample, observed
spectroscopically in an identical fashion, and thus forms an ideal
control sample.  The fact that it has a dramatically different
distribution of $q$ from the detached sample (the K-S probability that
they are drawn from the same distribution is 0.002), while expected
from binary evolution, reinforces that reality of the unexpected $q$
distribution for the detached sample.

\begin{figure}
\plotone{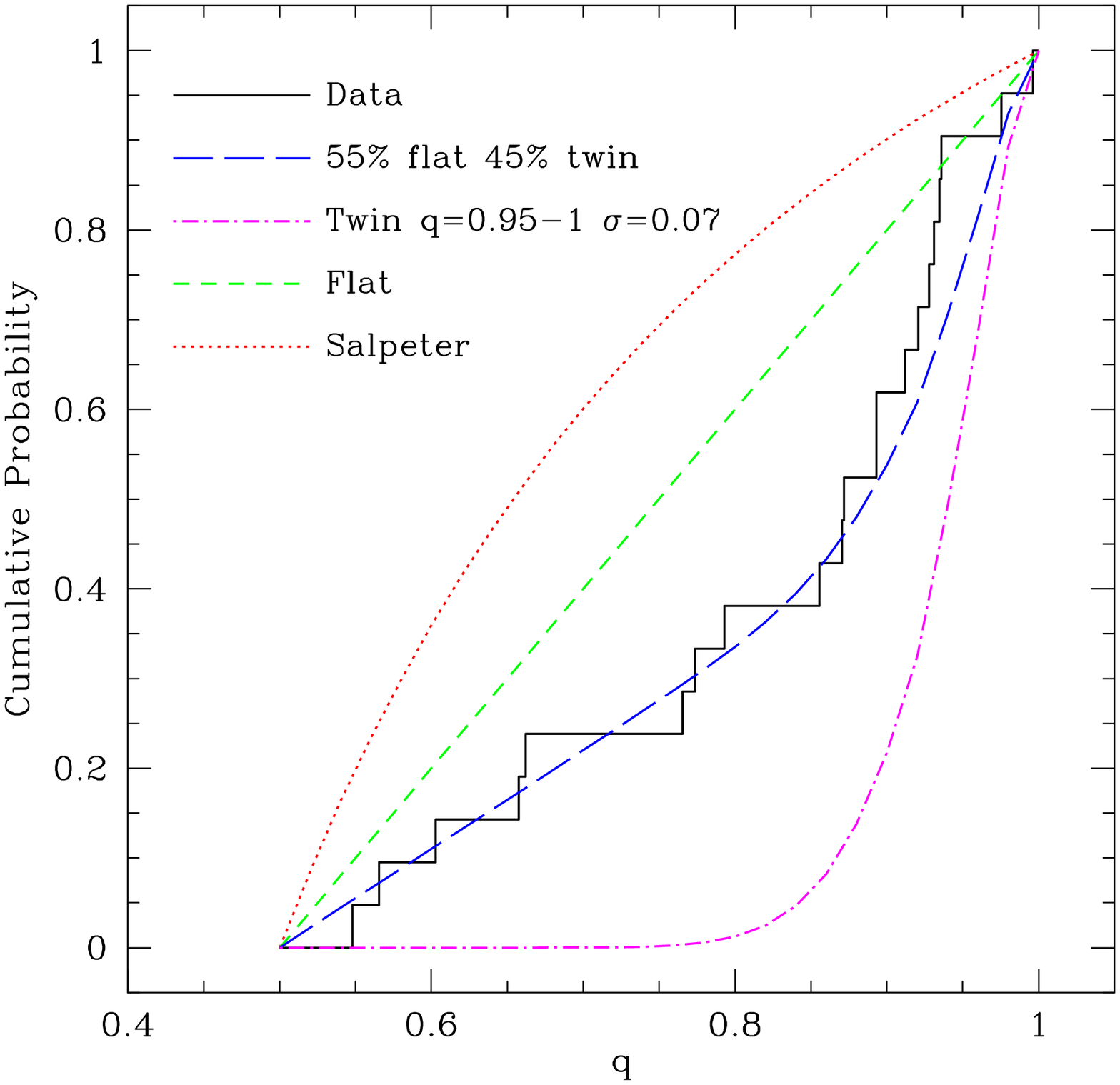}
\caption{The cumulative probability distribution of 21 SMC detached
eclipsing binaries as a function of the mass ratio $q=M_2/M_1$ are
compared with different model distributions.  The data (thick solid
line) are taken from Harries et al.  (2003) and Hilditch et
al. (2005).  The Salpeter relative IMF (dotted line) is clearly
incompatible, and a flat distribution (short-dashed line) is also
unlikely.  A fit including twins (45\% twins, 55\% flat) is given as
the long-dashed line, and the twin component alone ($q=0.95$ to 1
convolved with an observational error of 0.07) is indicated by the
dot-dashed line.}
\label{fig_ks}
\end{figure}

Given the above, we can now test statistically the various proposed
distributions of $q$ value in this interesting mass range.  For this
purpose we considered two different models for the distribution of
$q$: a flat mass distribution $p(q) = const$ and a Salpeter (1955)
relative mass function, $p(q)\propto q^{-2.35}$.  We assumed that
observational selection effects would cut off the $q$ distribution
below 0.5, so both the flat and Salpeter mass functions were truncated
at that value.  These distributions are illustrated in Fig.~2.  Both
the Salpeter and flat mass functions are inconsistent with the
observed data (probabilities of $ 3.4 \times 10^{-3}$ and 0.02,
respectively).  By varying the twin fraction we found that the best
fit was obtained with a flat component of 0.55 +/- 0.15 combined with
a twin distribution from q = 0.95 to 1 convolved with the
observational error $\sigma (q) = 0.07$; this is shown for comparison
in Fig.~2.  Based on this, we conclude that there is clear evidence
for two populations of eclipsing binaries, and that a substantial
fraction of them must be of nearly equal mass.  Even if we assume that
the absence of systems with mass ratios lower than $q=0.5$ is a
selection effect, we still require a significant twin population of
20-25\% with $q>0.95$ in the short period binary population (in
addition to the 5\% of such stars that would naturally be present in a
flat relative IMF).

\section{Other Evidence for a Significant Twin Population}

Because of the striking consequences of this result, it is reasonable
to ask how general it might be.  Fortunately, there has been an
dramatic increase in the quantity and quality of data in the last few
years.  Recent radial velocity surveys have provided large samples of
spectroscopic binaries.  Microlensing and planet eclipse studies have
provided a wealthy of information on eclipsing binaries.  In addition,
IR speckle studies in nearby open clusters have provided valuable
information on longer period systems.  The picture that emerges is
consistent with the idea that there are three natural components of
the relative binary mass function.  At long orbital periods ($P >
10\;$yrs) there is no evidence for a peak at q = 1 and the binary
masses appear to be uncorrelated.  However, for short period systems a
Salpeter relative IMF is ruled out in a number of different domains:
for example, field F and G stars (e.g. Halbwachs et al. 2003; Lucy \&
Ricco 1979); field halo stars (e.g. Goldberg, Mazeh \& Latham 2003);
massive stars (e.g. Garmany, Conti \& Massey 1980; this paper).

In addition to clear evidence for a flat component, a peak near $q =
1$ was reported by several investigators.  For example, Halbwachs et
al. (2003) studied a large sample of spectroscopic binaries type F7 to
K (masses from about 1.7 down to $0.5\;M_{\odot}$), including binaries
in open clusters (Pleiades and Praesepe).  They find that the mass
ratio has a broad peak from $q \approx 0.2$ to $q \approx 0.7$, and a
sharp peak for $q>0.8$ (what they call ``twins''). As they discuss,
the strength of the peak for $q>0.8$ gradually decreases with the
increasing orbital period, which is to be expected. In addition,
independent of period, the twin population has significantly lower
eccentricities compared to the other binaries, strongly suggesting a
different formation mechanism. The fraction of twins (see their Fig.9)
can be as high as 50\% for periods $P<10\;$days and it is still
significant (as high as 35\%) for much longer periods of up to 1000
days.

A much earlier study by Lucy \& Ricco (1979) also finds a strong and
narrow peak of binaries with $q\approx 0.97$, again using a sample of
spectroscopic binaries corrected for various observational errors and
biases. Tokovinin (2000) confirms that finding using additional modern
data and in fact also calls this population ``twins'', arguing that
they constitute 10-20\% of the total binary population in the
$P=2-30\;$days regime.

All these results most strongly suggest that the striking trend found
in this work for massive eclipsing SMC binaries is a general feature
rather than either a peculiar environmental effect or an observational
selection effect. In fact, it can be said that twins have been present
in most of such studies, even if not always recognized as such.

Additional, although perhaps more anecdotal support for the
significant twin population comes from the realms of very high, and
also very low mass stars found in eclipsing binaries. The most massive
binary known, WR 20a (Rauw et al. 2004; Bonanos et al.  2004),
contains $82.7\pm 5.5\;M_{\odot}$ and $81.9\pm 5.5\;M_{\odot}$
components (Rauw et al. 2005). The system happens to be an eclipsing
binary (Bonanos et al. 2004), so the masses of both components can be
measured accurately. The mass ratio for this system is $q=0.99\pm
0.05$, and it would be most interesting to measure this ratio to even
higher accuracy, to see if this system is indeed an ``identical
twin''. Given that $80\;M_{\odot}$ stars are extremely rare (both due to
the steepness of the mass function and their short lifetime), having
such a massive secondary would be most unlikely unless the twin
phenomenon is involved.

For very low stellar masses, several detached eclipsing binaries have
equally or even more strikingly close to $q=1$ values.  Lopez-Morales
\& Ribas (2005) report a new low-mass, double-lined, detached
eclipsing binary, GU Boo, with $M_1=0.610\pm 0.007\;M_{\odot},
M_2=0.599\pm 0.006\;M_{\odot}$, i.e. $q=0.98 \pm 0.02$. This is a very
similar system to the well-studied low mass eclipsing binary YY Gem,
which as shown by Torres and Ribas (2002) has two components ``which
are virtually identical to each other", with a mass of $M_{1,2}=0.5992
\pm 0.0047\;M_{\odot}$ and $q=1.0056 \pm 0.0050$, a truly ``identical
twin''.  Even lower mass detached eclipsing system, CM Dra, has a mass
ratio of $q=0.9260 \pm 0.0026$, and $M_1=0.231\;M_{\odot}$ (Metcalfe
et al. 1996), again very close to $q=1$.  It should be said that there
are very few such law-mass detached eclipsing binaries known, making
the very close values of ``q'' described above even more striking.

While we most strongly argue for the existence of a significant, at
least 20\% and maybe as high as 50\% twin population, based on the
work in this paper and a vast body of work described in the
literature, additional studies of the twin population would be most
useful. For example, as mentioned earlier in this paper, the sample of
50 SMC binaries studied by HHH03 and HH05 comes from a much larger
sample of 169 eclipsing binaries they observed spectroscopically. Even
if the RV solutions for the remained of the sample are not good enough
to study their detailed properties such as mass, it could possibly be
used to study the distribution of $q$, especially if additional RV
epochs could be obtained.  There is also a large OGLE catalog of 2580
eclipsing binaries in the LMC (Wyrzykowski et al. 2003), which could
be used to select bright targets for analogous RV study in that
galaxy.

\section{Discussion}

\subsection{Implications for the Binary Neutron Star Birthrate}

The first application of our result is to the inferred production rate
of binary neutron stars (NS+NS) and neutron star + black hole (NS+BH)
systems.  The production of double compact systems is a result of
close binary interactions.  The primary will experience a supernova
first if $M_p > M_{crit}(NS)$.  It will leave a black hole if $ M_p >
M_{crit} (BH)$ and a neutron star otherwise. However, Bethe, Brown,
and Lee (2005a,b; hereafter BBL05) argue that mass transfer from the
envelope of the secondary onto the compact object will lead to strong
accretion, resulting in a black hole remnant from the primary unless
the secondary is very similar in lifetime and mass to the primary.
Furthermore, they adopt a steep mass function for the mass of the
secondary relative to the primary (Salpeter 1955), which makes such a
near equality in mass unlikely.  They therefore argue that there
should be a large population of NS+BH systems relative to the NS+NS
population.  Any given measurement of the NS+NS population therefore
would imply the presence of a much larger NS+BH population.  Although
the latter population would be difficult to detect directly, it would
produce a large signal in gravitational wave experiments such as LIGO.

BBL05 adopted $M_{crit}(NS) = 10\;M_{\odot}$ and $M_{crit}(BH) =
20\;M_{\odot}$, requiring masses equal to within 4\% to produce NS+NS
systems and otherwise producing NS+BH systems.  They also assumed both
the primary and secondary mass functions were slightly steeper than
the Salpeter IMF, namely $dn/dm \propto m^{-2.5}$, and furthermore
that the secondary IMF was truncated at 10$\;M_{\odot}$.  From this
set of assumptions they estimate a NS+BH coalescence rate ten times
higher than the NS+NS rate and a detectability of the former twice as
high, resulting in a factor of 20 boost in the predicted LIGO rate.

A very different picture emerges from a binary population including
stellar twins.  In order to estimate the impact of the relative IMF on
the double compact object production rate, we proceed as follows.  We
assume, as per BBL05, that all systems with both primary and secondary
masses between 10 and 20$\;M_{\odot}$ will produce double compact
object systems; furthermore, we assume that systems where the mass
difference is greater than 5\% may produce NS+BH systems.  We also
adopt a Salpeter primary IMF, $dn/dm\propto m^{-2.35}$.  We consider
three mass function cases: a flat mass function ($dn/dm = const$); a
population with 25\% twins and 75\% drawn from a flat mass function;
and a Salpeter relative secondary IMF.

For a flat mass function 24\% of the close binaries with primaries in
the relevant mass range could produce double compact systems, and 5\%
of them have masses within 5\%.  In this case the maximum ratio of
(NS+BH)/(NS+NS) systems is 5.  The same calculations for a relative
Salpeter IMF with a minimum mass of $ 0.3 M_{sun}$ yields 0.32\% and
0.044\% respectively, for a ratio of 6 (but a very small parent
population).  However, the twin population would imply that 43\% of
close binaries could produce double compact systems and 30\% would be
nearly equal in mass, for an inverted ratio of (NS+BH)/(NS+NS) of 0.3.
We therefore conclude that there is no requirement for a large unseen
population of NS+BH systems, even before efficiency arguments are
accounted for, and that it is therefore unlikely that the LIGO event
rate should be substantially amplified by such systems.  Population
synthesis models (e.g. Belczynski, Bulik \& Ruiter 2005) predict
(NS+BH)/(NS+NS) rates of less than 1/3 even before twins are accounted
for, so that the actual ratio may well be much less than the simple
estimates above.

\subsection{Implications for the SNIa Rate}

The presence of binary twins could have intriguing consequences for
the time delay prior to the onset of Type Ia supernovae.  The Type Ia
SN rate has been measured out to z=1.6 by the GOODs consortium, and
the peak in the observed rate occurs at a significantly lower redshift
than the peak in the star formation rate.  This led Strolger et
al. (2004) to infer a long delay time, of order 2-5 Gyr depending upon
the assumed functional form for the delay time; a similar timescale of
1.7 Gyr was obtained by Gal-Yam \& Maoz (2004).  However, studies of
the type Ia SN rate in the local universe indicate a strong
correlation between the star formation rate and the Type Ia SN rate
(Mannucci et al. 2005a).  Scannapieco \& Bildsten (2005) and Mannucci,
Della Valle, \& Panagia (2005b) argue that this data requires a
two-component model for the SNIa rate: a prompt component with a short
delay time ($\sim0.1\;$Gyr) and a delayed component with a timescale
comparable to that observed at high redshift ($>1.7\;$Gyr).  The
existence of a prompt SNIa component is not inconsistent with Galactic
chemical evolution data because it would be correlated with the SNII
rate (see Scannapieco \& Bildsten 2005).  In such models the decrease
in [O/Fe] with increased [Fe/H] is caused by the increase in the SNIa
rate from the onset of the delayed component rather than from the
absence of SNIa at early times.

There are two general formation scenarios for type Ia SN.  A white
dwarf could accumulate mass from a non-degenerate companion star
(e.g. Whelan \& Iben 1973), or two white dwarfs could merge
(e.g. Webbink 1984).  Following Belczynski et al. (2005) the delay
timescale has two main components. Both the primary and secondary must
leave the main sequence; this timescale is governed by the lifetime of
the lower mass secondary.  In the case of binary WD mergers the
timescale for coalescence is also important, and it is primarily a
function of the orbital separation after the envelopes are removed.
The degree of orbital shrinkage in turn depends on the detailed
physics of the common envelope phase.  We argue that the presence of a
substantial fraction of binary twins will have a substantial impact on
both of these timescales.  The secondary in twin systems will have a
lifetime similar to the primary; as a result, the evolutionary delay
time will be minimized.  A division into two relative mass functions
(twin and flat) will naturally lead to a comparable division into
evolutionary timescales.  However, as we will see below, there are
also distinctive features of twin common envelope evolution that could
play an addition role.

\subsection{Double WD Systems and Common Envelope Evolution}

Double white dwarf systems are thought to arise from systems which
have experienced two common envelope events, and the white dwarf
masses are primarily determined by the orbital separation at the onset
of the common envelope phase.  Recent observational studies have found
that such systems tend to have very similar masses (Maxted, Marsh \&
Moran 2002).  The most natural implication is that the orbital period
of the progenitor system must not have changed much during the first
common envelope event (to permit similar white dwarf masses), but that
it also must have changed dramatically during the second event (to
permit the short final orbital periods.  Such a configuration is
expected to be rare in traditional population synthesis calculations
(e.g. Iben, Tutukov \& Yungelson 1997).  However, Nelemans et
al. (2000) and Nelemans \& Tout (2005) argue that in their model of
common envelope evolution stars of nearly equal initial mass will
experience relatively little orbital shrinkage in the first mass
transfer event.  Although their treatment of the CE phase can
reconstruct the observed distribution of short period white dwarf
binaries, the absolute rates predicted with a flat IMF are on the low
end.  The inclusion of binary twins in the relative IMF would solve
this potential problem by boosting the production rate.  By extension,
the inclusion of binary twins could also make it easier to produce
pairs of more massive white dwarfs for doubly degenerate SN Ia
scenarios.

\subsection{Blue Stragglers}

Finally, we briefly note the consequences for the production of blue
stragglers through binary mergers.  Although blue stragglers can be
produced by a variety of mechanisms, one mechanism that is likely to
be important is coalescence of a close main sequence binary through
angular momentum loss.  Both the merger rate and whether a merger is
observable as a blue straggler are sensitive to the assumed relative
binary IMF (Andronov, Pinsonneault \& Terndrup 2005).  The inclusion
of twins would boost the merger rate because, for old systems, more
massive secondaries lose more angular momentum.  It would also make
the merger products more massive on average and more likely to be
detected as blue stragglers observationally.

\section{Possible Relevance for Other Studies}

Given the origin of this paper (see Acknowledgments), instead of the
usual ``Summary'' we have decided to provide a (naturally, incomplete)
list of recent astro-ph postings for which the ``binaries like to be
twins'' result is more or less relevant. In no particular order:

\begin{enumerate}

\item The equal masses of the neutron stars observed in the binary
NS-NS systems would follow naturally from their evolution from the
binary twins system, and probably does not require special tuning,
unlike suggested in BBL05 papers.  This could be relevant to Piran \&
Shaviv (2005).

\item In light of recent localizations of short gamma-ray bursts,
NS-NS and NS-BH binaries were discussed in a whole slew of recent
astro-ph postings (e.g. Villasenot et al. 2005; Fox et al.  2005;
Hjorth et al. 2005). As discussed, having a large population of
equal-mass binaries directly affects the predicted rates of such
events.

\item Another recent GRB event which resulted in a slew of astro-ph
postings was the discovery of the $z=6.3$ GRB 050904 (e.g. Haisplip et
al. 2005; Price et al. 2005; Tagliaferri et al. 2005). That event was
seen as opening the epoch of studying GRBs at even higher redshifts,
with possible relevance to Population III stars. While GRB 050904 was
a standard ``long-soft'' GRB, and therefore most likely originating
from a core-collapse ``hypernova'' (e.g. Stanek et al. 2003),
Population III stars, lacking metals, might be hard pressed to lose
their envelopes (e.g.  Kudritzki 2002), so binaries might be needed to
do the trick (e.g. Bromm \& Loeb 2005).

\item In a more local Universe, Stepien (2005) discussed a possible
solution to the ``Kuiper paradox'' seen in the binary W UMa-type
systems. A large population of equal-mass binaries would be certainly
relevant, although possibly not crucial, to his proposed scenario.

\item Krumholz, McKee and Klein (2005) argue that ``Stars Form By
Gravitational Collapse, Not Competitive Accretion'' (but see e.g. Bate
2000). It would be most interesting to see if the existence of large
population of twins favors any of the existing star formation models
and in fact Bonnell \& Bate (2005) claim exactly that.

\end{enumerate}

\acknowledgments

We would like to thank the participants of the morning ``Astronomy
Coffee'' at the Department of Astronomy, The Ohio State University,
for the daily and lively astro-ph discussion, one of which has
prompted us to investigate the issue of the mass ratio in massive
binaries. We especially thank John Beacom and Andy Gould for their
useful input.  KZS thanks Alceste Bonanos for many useful discussions
on the most massive stars. We also thank Alceste Bonanos, Jonathan
Devor, Andy Gould, Joel Hartman and Bohdan Paczy\'nski for their
comments on earlier versions of this draft.

\end{document}